\documentclass[sigconf]{acmart}
\newcommand{\modelname}{\textsf{PMTRec}\xspace}
\usepackage{algorithm}
\usepackage{algorithmic}

\AtBeginDocument{%
  \providecommand\BibTeX{{%
    \normalfont B\kern-0.5em{\scshape i\kern-0.25em b}\kern-0.8em\TeX}}}

\setcopyright{acmcopyright}
\copyrightyear{2018}
\acmYear{2018}
\acmDOI{XXXXXXX.XXXXXXX}

\acmBooktitle{Woodstock '18: ACM Symposium on Neural Gaze Detection,
 June 03--05, 2018, Woodstock, NY} 
\acmPrice{15.00}
\acmISBN{978-1-4503-XXXX-X/18/06}
\usepackage{multirow}
\usepackage{xspace}
\usepackage{color}
\usepackage{amsmath}
\usepackage{enumitem}

\begin{document}

\title{Personalized Multi-task Training for Recommender System}
\author{Liangwei Yang}
\authornote{The work is finished during an internship at Salesforce AI Research}
\affiliation{%
  \institution{University of Illinois Chicago}
  \city{Chicago}
  \country{USA}}
\email{lyang84@uic.edu}

\author{Zhiwei Liu}
\affiliation{%
  \institution{Salesforce AI Research}
  \city{Palo Alto}
  \country{USA}}
\email{zhiweiliu@salesforce.com}

\author{Jianguo Zhang}
\affiliation{%
  \institution{Salesforce AI Research}
  \city{Palo Alto}
  \country{USA}}
\email{jianguozhang@salesforce.com}

\author{Rithesh Murthy}
\affiliation{%
  \institution{Salesforce AI Research}
  \city{Palo Alto}
  \country{USA}}
\email{rithesh.murthy@salesforce.com}

\author{Shelby Heinecke}
\affiliation{%
  \institution{Salesforce AI Research}
  \city{Palo Alto}
  \country{USA}}
\email{shelby.heinecke@salesforce.com}

\author{Huan Wang}
\affiliation{%
  \institution{Salesforce AI Research}
  \city{Palo Alto}
  \country{USA}}
\email{huan.wang@salesforce.com}

\author{Caiming Xiong}
\affiliation{%
  \institution{Salesforce AI Research}
  \city{Palo Alto}
  \country{USA}}
\email{cxiong@salesforce.com}

\author{Philip S. Yu}
\affiliation{%
  \institution{University of Illinois Chicago}
  \city{Chicago}
  \country{USA}}
\email{psyu@uic.edu}

\renewcommand{\shortauthors}{Liangwei Yang et al.}

\begin{abstract}
In the vast landscape of internet information, recommender systems (RecSys) have become essential for guiding users through a sea of choices aligned with their preferences. These systems have applications in diverse domains, such as news feeds, game suggestions, and shopping recommendations. Personalization is a key technique in RecSys, where modern methods leverage representation learning to encode user/item interactions into embeddings, forming the foundation for personalized recommendations. However, integrating information from multiple sources to enhance recommendation performance remains challenging.
This paper introduces a novel approach named \modelname, the first personalized multi-task learning algorithm to obtain comprehensive user/item embeddings from various information sources. Addressing challenges specific to personalized RecSys, we develop modules to handle personalized task weights, diverse task orientations, and variations in gradient magnitudes across tasks. \modelname dynamically adjusts task weights based on gradient norms for each user/item, employs a Task Focusing module to align gradient combinations with the main recommendation task, and uses a Gradient Magnitude Balancing module to ensure balanced training across tasks.
Through extensive experiments on three real-world datasets with different scales, we demonstrate that \modelname significantly outperforms existing multi-task learning methods, showcasing its effectiveness in achieving enhanced recommendation accuracy by leveraging multiple tasks simultaneously. Our contributions open new avenues for advancing personalized multi-task training in recommender systems. 
\end{abstract}




\maketitle

\section{Introduction}
Regarding the pervasive landscape of information available on the Internet~\cite{mayer2013big}, recommender systems (RecSys)~\cite{lu2012recommender} play a pivotal role in facilitating the exploration of items for users.
RecSys improves the efficiency of information digestion 
and has permeated into every corner of our lives, such as news feeds~\cite{wu2022feedrec}, game suggestions~\cite{yang2022large}, and shopping recommendations~\cite{gu2020hierarchical}. The great commercial value of RecSys also establishes its indispensability in web services and drives the success of many Internet companies~\cite{gomez2015netflix,ying2018graph,liu2022monolith}.

Personalization is the key technique for an effective RecSys~\cite{qian2013personalized}. 
Personalized RecSys suggests specific candidate items for each user based on his/her own historical interactions. It offers each user a distinct experience with the web service tailored to their interests. Due to the success of representation learning~\cite{bengio2013representation}, current state-of-the-art methods~\cite{he2020lightgcn,wang2022towards} in RecSys encode user/item interactions into dense vector representations, referred to as embeddings. The embedding of each user/item contains his/her/its unique information, which is the bedrock for providing personalized recommendations. Based on the encoded embedding, these methods recommend items with the highest similarities for each user. Besides user-item interactions, more information can be found in web services and applications such as users' social networks~\cite{yang2021consisrec} and item features~\cite{he2016ups}. Then, a natural question arises: Can we obtain a more comprehensive embedding from multiple sources of information to improve the recommendation performance?

\begin{figure}[bp]
    \centering
    \includegraphics[scale=0.5]{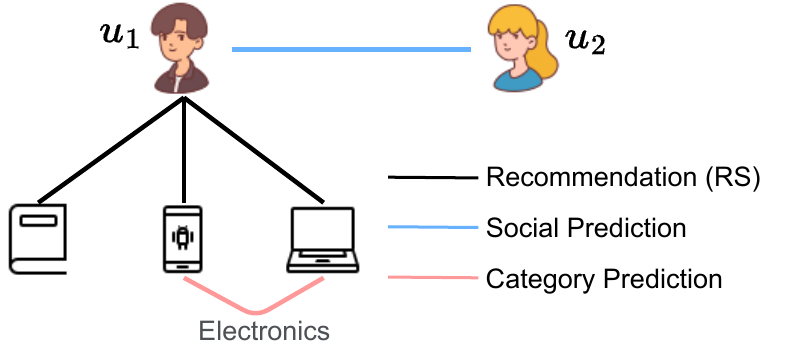}
    \caption{A toy example of multi-task learning in RecSys.}
    \label{fig:illustration}
\end{figure}

In this paper, we seek to answer this question with multi-task learning~\cite{zhang2018overview}, which has been shown effective in various application domains such as computer vision~\cite{yu2020gradient} and natural language processing~\cite{dong2015multi}. The core idea behind it is to obtain a comprehensive user/item embedding that can perform multiple tasks at the same time. As illustrated in Figure~\ref{fig:illustration}, we show three tasks associated with distinct information. 
The black edges denote the recommendation task to predict users' preference for items, which is the main task for optimization. 
The blue edge indicates social information, where the social link prediction task between users characterizes social information into embedding of users. 
The red line represents two items in the same ``Electronics'' category. 
Predicting whether two items share the same category encodes item category similarity into item embedding. 
Ideally, training user/item embedding with all these tasks endows incorporating information from multiple sources, leading to a more comprehensive vector representation.

Several key challenges are still under-explored when developing a multi-task learning framework for personalized RecSys, which are threefold as follows: 
(1) Personalized task weights. For each user, the importance of different tasks should be distinct.
As illustrated in Figure~\ref{fig:illustration}, the suitable task weights for $u_1$ and $u_2$ are different.
$u_1$ has sufficient item interactions to obtain an accurate gradient for updating embedding, which suggests a higher RecSys task weight for $u_1$. However, $u_2$ does not interact with any item before, in which case the RecSys task fails to provide informative information, and it is beneficial for $u_2$ to have a higher weight on the Social Prediction task. Supporting specific task weights for each user is the first challenge, and it is essential for personalized training with multiple tasks.
(2) Varied task orientations. The purpose of different tasks is varied, leading to diverged embedding updates during training, which may conflict with each other or even be meaningless for improving the main RecSys task. 
Handling the varied updating orientations from multiple tasks to improve the RecSys task poses the second challenge.
(3) Difference in gradient magnitude across tasks. The loss functions for multiple tasks are inconsistent, and thus the gradient magnitude could be magnificently different. 
Directly averaging all those gradients from multiple tasks may overlook tasks with smaller gradients.
However, balancing the gradient magnitude difference and making the small gradient counts are an important yet under-explored challenge.

This paper proposes the first Personalized Multi-task Training algorithm for Recommender System (\modelname) to obtain a more comprehensive user/item embedding from multiple information sources. Different modules are designed accordingly to deal with specific challenges. 
We dive into the gradient level during the backward propagation to enable personalized task weights. \modelname first collects the gradient from each task during backpropagation and then combines the gradient on each user embedding separately. \modelname conducts meticulous user/item-specific gradient level operations rather than the existing task-wise algorithms~\cite{yu2020gradient,wang2020gradient,chen2018gradnorm,lin2021reasonable}.
For a specific gradient of one user from one task, \modelname firstly calculates the norm to measure the importance. 
A gradient with a higher magnitude signifies a greater impact on optimizing the corresponding task. 
\modelname then utilizes the gradient norm to calculate task weight for gradient combination. The gradient norm on each task for different users is distinct, leading to personalized gradient weights.
We also designed a \textit{Task Focusing} module to tackle the varied task orientation challenge. 
We enforce the learning process gradually, focusing on the main RecSys task during the training stage, achieved by increasing the corresponding gradient combination weight. 
After initial exploration of other tasks, the Task Focusing module leads to the combined gradient gradually aligning well with the RecSys task, leaving the varied task orientations imperceptible at the final stage.
Different tasks generate gradients of different magnitudes. Multi-task training algorithms overlook gradients with smaller magnitudes without considering the difference.
Thus, we further propose the Gradient Magnitude Balancing module to deal with this challenge and balance the training of all tasks. Our contributions can be summarised as follows:
\begin{itemize}[leftmargin=*]
    \item Conceptually, for personalized multi-task learning, we first illustrate the problem, identify the challenges, and propose feasible solutions for recommender system.
    \item Methodologically, we propose the novel \modelname algorithm that operates on the gradient level during backpropagation to provide personalized gradient weights.
    \item Experimentally, we conduct extensive experiments on three real-world datasets with varied scales to justify the effectiveness of \modelname compared with current multi-task training algorithms.
\end{itemize}

\section{Preliminaries}
This section illustrates the preliminaries for \modelname, including a problem statement and multi-task learning.
\subsection{Problem Statement}
Given a set of users $\mathcal{U} = \{u_1,u_2,...,u_{\left | \mathcal{U} \right|}\}$ and a set of items $\mathcal{I} = \{i_1,i_2,...,i_{\left | \mathcal{I} \right|}\}$, the personalized RecSys task aims to generate a list of items for a given user $u$.
The most important information for training a recommendation model is the currently observed historical interactions, which can be represented as a user-item bipartite graph $\mathcal{G}=(\mathcal{V}, \mathcal{E})$, where $\mathcal{V}= \mathcal{U}\cup \mathcal{I}$ and there is an edge $(u,i) \in \mathcal{E}$ between $u$ and $i$ if $u$ has interacted with $i$ historically. The adjacency matrix of $\mathcal{G}$ is represented by $\textbf{R}\in \mathbb{R}^{\left|\mathcal{V}\right| \times \left|\mathcal{V}\right|}$. For implicit feedback~\cite{rendle2012bpr} such as clicks, views, and purchases, $R_{u,i}=1$ if user $u$ has interacted with item $i$, and $R_{u,i}=0$ otherwise. For explicit feedback, the user provides, such as rating score, $R_{u,i}$ is the explicit rating score from $u$ to $i$. Besides the direct historical interactions, auxiliary information can also be observed from both the user and item side, which is termed by profile $\mathcal{P}=(\mathcal{P}^u,\mathcal{P}^i)$. $\mathcal{P}^u$ contains all the auxiliary information from the user side, such as features and social friends. $\mathcal{P}^i$ represents the item side auxiliary information such as category and co-view information. Based on all the above information, personalized RecSys aims to predict the preference score in the adjacency matrix of $\mathcal{G}$ and recommends a list of $k$ items with the top-k highest score.

\subsection{Multi-task Learning}
Multi-task learning (MTL) aims to learn a comprehensive model to accomplish multiple tasks by training them simultaneously. This learning paradigm assumes different tasks can enhance each other and perform better on all tasks. Without loss of generality, we integrate and illustrate the current multi-task learning algorithms~\cite{yu2020gradient,wang2020gradient,chen2018gradnorm,lin2021reasonable} from the gradient level. Each task is trained and can be categorized by a loss function $\mathcal{L}$, which directly leads to a gradient $\bigtriangledown \mathcal{L}$ after backpropagation~\cite{rumelhart1986learning}. Assuming we have $T$ tasks, then a list of the gradient for different tasks is obtained $(\bigtriangledown\mathcal{L}_1, \bigtriangledown\mathcal{L}_2, ..., \bigtriangledown\mathcal{L}_T)$. An MTL algorithm can be seen as a pooling function over the gradient list to obtain the combined gradient for descent $\bigtriangledown=\text{Pool}_{\text{MTL}}(\bigtriangledown\mathcal{L}_1, \bigtriangledown\mathcal{L}_2, ..., \bigtriangledown\mathcal{L}_T)$. Different MTL algorithms differ in the design of the pooling function, which can be parameterized or heuristic.

\section{Method}
\begin{figure*}[!hbt]
    \centering
    \includegraphics[width=1.1\linewidth]{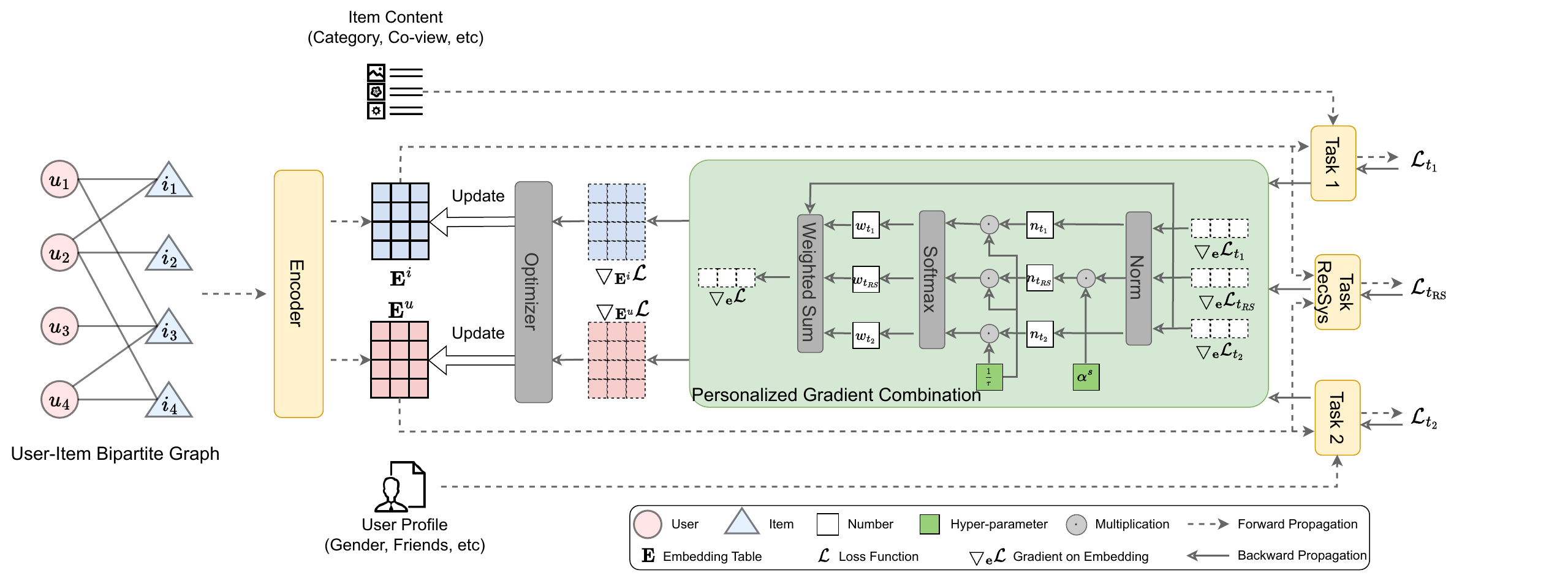}
    \caption{Model Framework. Our proposed \modelname is illustrated in detail with two propagation stages. (1) Forward propagation computes the user/item embedding and the losses for different tasks. (2) Backward propagation pools the gradient obtained from each task with a personalized gradient combination for each user/item. The combined gradient is then used to update the embedding with an optimizer. (Best viewed in color)}
    \label{framework}
\end{figure*}

This section demonstrates the proposed \modelname, shown in Figure~\ref{framework}. It includes a forward propagation stage to gather gradients from multiple tasks and a backward stage to aggregate gradients from different tasks individually. Our essential contribution lies in the backward stage, which contains the task focusing and gradient magnitude balancing modules.

\subsection{Forward Propagation}
In the forward propagation stage, \modelname computes the losses for different tasks. It consists of an encoder to obtain the user/item representation and several task modules for loss computation.

\subsubsection{Encoder}\label{sec:encoder}
Current state-of-the-art personalized RecSys~\cite{he2020lightgcn,wang2022towards} learns an encoder to embed historical interactions $\mathcal{G}$ into dense embedding representations as:
\begin{equation}
    \mathbf{E} = \text{Encoder}(\mathcal{G}),
\end{equation}
where $\mathbf{E}=(\mathbf{E}^u, \mathbf{E}^i)$. $\mathbf{E}^u\in \mathbb{R}^{|\mathcal{U}|\times d}$ and $\mathbf{E}^i \in \mathbb{R}^{|\mathcal{I}|\times d}$ represent the user/item embedding separately. 
$\mathbf{e}_u\in \mathbf{E}^u$ is user $u$'s embedding, which encodes the specific information related to $u$'s historical interaction. $\mathbf{e}_u$ is the bedrock to provide personalized recommendation for $u$, and $d$ is the dimension size. In \modelname, we use matrix factorization~\cite{rendle2012bpr} as the encoder to directly embed interactions into dense user/item vectors. \modelname aims to encode more information into those vector representation with auxiliary tasks.

\subsubsection{Multi-task Loss computation}
For RecSys, tasks $\mathcal{T}$ are classified into the recommendation task $t_{RecSys}$ and auxiliary tasks $\mathcal{T}_{\text{Aux}} = \{t_1, t_2, ..., t_{|\mathcal{T}_{\text{Aux}}|}\}$. Auxiliary tasks can be built in different ways based on the available information. As illustrated in Figure~\ref{framework}, Task 1 is built purely from item embedding $\mathbf{E}^i$ and item profile $\mathcal{P}^i$. It aims to embed the item's profile information to the corresponding item embedding. Task 2 is built in a similar way from the user side. Tasks should be designed based on the available information for each dataset. For example, we demonstrate the task design in detail for the three experimental datasets in Section~\ref{sec:experiment}. Feeding the embedding table $\mathbf{E}$ encoded from Section~\ref{sec:encoder} into the loss computation module of different tasks, we can obtain a list of losses:
\begin{equation}\label{eq:losses}
    (\mathcal{L}_{RecSys}, \mathcal{L}_1, \mathcal{L}_2, ...,\mathcal{L}_{|\mathcal{L}_{\text{Aux}}|})
\end{equation}

\subsection{Backward Propagation}
During the backward propagation stage, \modelname combines the gradients in a personalized fashion, i.e., computes specific gradient weights for each user. The backward stage consists of gradient collection, task focusing and gradient magnitude balancing modules.

\subsubsection{Gradient Collection}\label{sec:grad_collect}
For each user/item, each loss $\mathcal{L}$ in Equation~\ref{eq:losses} leads to one gradient on the corresponding embedding vector, denoted as $\bigtriangledown_{\mathbf{e}}\mathcal{L}$. Then a collection of gradients from different tasks on the specific user/item is represented by:
\begin{equation}\label{eq:grads}
    (\bigtriangledown_{\mathbf{e}}\mathcal{L}_{RecSys}, \bigtriangledown_{\mathbf{e}}\mathcal{L}_1, \bigtriangledown_{\mathbf{e}}\mathcal{L}_2, ...,\bigtriangledown_{\mathbf{e}}\mathcal{L}_{|\mathcal{L}_{\text{Aux}}|}),
\end{equation}
where $\mathbf{e}$ is the embedding for one user/item.
Then \modelname measures the impactfulness of the gradient by $l_2$ norm as $n_{t}=|\bigtriangledown_{\mathbf{e}}\mathcal{L}_{t}|_2$. A larger $n_{t}$ indicates the current user/item is more impactful to optimize task $t$. Then a list of norms is computed from Equation~\ref{eq:grads} as $(n_{RecSys}, n_{1}, n_{2}, ..., n_{|\mathcal{L}_{\text{Aux}}|})$.

\begin{algorithm}[t]
  \caption{The training process of \modelname}
  \label{algo}
  \begin{algorithmic}[1]
  \REQUIRE Training set~$\mathcal{G}$, user/item embedding table $\mathbf{E}$, main RecSys task $\mathcal{L}_{\text{RecSys}}$, auxiliary tasks $\mathcal{L}_{\text{Aux}}=\{\mathcal{L}_1, \mathcal{L}_2, ..., \mathcal{L}_{|\mathcal{\mathcal{L}_{\text{Aux}}}}|\}$, a hyper-parameter $\tau$ to control the gradient magnitude, total training epoch $S$, learning rate $lr$, task-focusing base $\alpha$.
  \FOR{$s = 1,2,3, \dots ,$ to $S$}
  \FOR{Each mini-batch of user-item interactions $\mathcal{B}$}

  \STATE $Grad$ = [$\bigtriangledown_{\mathbf{E}}\mathcal{L}_{\text{RecSys}}(\mathcal{B})$]
  
  {\color{cyan}{\textit{Gradient Collection}}}

  \FOR{$\mathcal{L}$ $\in$ $\mathcal{L}_{\text{Aux}}$}
  \STATE  $Grad_{\mathcal{L}}=\bigtriangledown_{\mathbf{E}}\mathcal{L}(\mathcal{B})$
  
  \STATE  $Grad$ append $Grad_{\mathcal{L}}$
  
  \ENDFOR
  

  \STATE $Grads$ = Stack($Grad$, dim = 0)

  \STATE $Norm$ = Norm($Grads$, dim = -1)
  
  {\color{cyan}{\textit{Task Focusing}}}

  \STATE $Norm[0]$ = $Norm[0] * \alpha^{s}$
  
  {\color{cyan}{\textit{Gradient Magnitude Balancing}}}
  
  \STATE $Attn_{weight}$ = Softmax($Norm / \tau$, dim = 0)

  \STATE $\bigtriangledown_{\mathbf{E}}$ = Sum($Attn_{\text{weight}}$.unsqueeze(-1) * $Grads$, dim = 0).

  \STATE $\mathbf{E} = \mathbf{E} - lr * \bigtriangledown_{\mathbf{E}}$.
  
  \ENDFOR
  \ENDFOR
  \end{algorithmic}
\end{algorithm}

\subsubsection{Task Focusing}\label{sec:focus}
For personalized multi-task training, \modelname treats the RecSys task as the main task and all the other tasks as auxiliary ones for further improvement. Auxiliary tasks are built differently, leading to varied task orientations and may even contradict the main task. Thus we design the Task Focusing module to focus the training on the main RecSys task gradually. To this end, a hyper-parameter $\alpha>1$ is introduced as the increasing base. Before transforming the gradient norm from Section~\ref{sec:grad_collect} into gradient weights, we multiply the weight on the main RecSys task as:
\begin{equation}\label{eq:alpha}
    n_{RecSys} *= n_{RecSys} * \alpha^{s},
\end{equation}
where $s$ is the current epoch number. In this design, \modelname will explore all the tasks first and gradually focus the training on the main RecSys task. The training will be aligned with the main RecSys task at the final stage. A larger $\alpha$ will fasten the focusing speed. 
Though in exponential design, the following Softmax computation with built-in log-sum-exp trick ensures the number stability.

\subsubsection{Gradient Magnitude Balancing}\label{sec:balance}
The loss functions within Equation~\ref{eq:losses} are different due to the varied task designs, making the gradient magnitude essentially different (more than 10 times). The gradient with a small magnitude will be overlooked if we directly transform the gradient norm into gradient weight. Thus, we propose the gradient magnitude balancing module to balance the impact of different gradient magnitudes. In detail, the gradient weight is obtained by a Softmax with temperature:
\begin{equation}\label{eq:softmax}
    w_{t} = \frac{\exp^{n_t / \tau}}{\sum_{t^*\in \mathcal{T}}\exp^{n_{t^*} / \tau}},
\end{equation}
where $\tau$ is the hyper-parameter to balance gradient magnitude. For example, all task weights will be the same without being impacted by different gradient magnitudes when $\tau \to +\infty$. Conversely, \modelname will only count the gradient with the largest norm when $\tau \to -\infty$. A suitable $\tau$ balance should be selected for the best gradient magnitude. With Equation~\ref{eq:softmax}, we can obtain the gradient weights from different tasks as $(w_{RecSys}, w_{1}, w_{2}, w_{|\mathcal{L}_{\text{Aux}}|})$. Then \modelname obtains the gradient on the corresponding user/item by weighted sum on the gradients:
\begin{equation}\label{eq:softmax}
\bigtriangledown_{\mathbf{e}}\mathcal{L}=\sum_{t\in\mathcal{T}}w_{t}\bigtriangledown_{\mathbf{e}}\mathcal{L}_{t},
\end{equation}
where $\bigtriangledown_{\mathbf{e}}$ is the combined gradient for a specific use/item and is used to update corresponding embedding $\mathbf{e}$. The variance in gradient norms observed across distinct tasks for different users/items culminates in the derivation of personalized gradient combination weights. This distinctive attribute positions \modelname as a pioneering instance of personalized multi-task training algorithms.

\subsection{Discussion}
RS naturally faces the data sparsity issue~\cite{yang2023graph,yin2020overcoming} as the candidate item pool is too large for each user to explore. Information from external knowledge such as social networks~\cite{yang2021consisrec} and knowledge graph~\cite{wang2019kgat} has proven to provide extra information to enrich user/item representation. However, previous methods model external knowledge as part of the encoder, such as adding auxiliary edges besides the user-item bipartite graph, which needs sophisticated design within the encoder. \modelname proposes another solution, i.e., embed the external knowledge into user/item representation with multi-task training. It can keep the model intact and embed that information with more tasks. 

We further analyze the overhead time complexity caused by multi-task training. For each user/item, \modelname firstly computes the norm of an embedding gradient in Section~\ref{sec:grad_collect}, which takes ($O(d)$). The Task Focusing module in Section~\ref{sec:focus} will only cost $O(1)$ as it only computes on the RecSys task. Then the Gradient Magnitude Balancing module in Section~\ref{sec:balance} takes $O(|\mathcal{T}|)$. To sum up, the overall time complexity of \modelname is $O((|\mathcal{U}| + |\mathcal{I}|)\times(d + |\mathcal{T}|))$, where $\mathcal{U}, \mathcal{I}, \mathcal{T}$ is the user/item/task set respectively and $d$ is the embedding dimension. It is noted that the $(|\mathcal{U}| + |\mathcal{I}|)$ dimension can be easily batched up for parallel computation, which makes the wall clock time even much faster. The batched version of \modelname is shown in Algorithm~\ref{algo} with all the modules marked correspondingly, and $\bigtriangledown_{\mathbf{E}}$ is the aggregated gradient on embedding $\mathbf{E}$.

\section{Experiment}\label{sec:experiment}
This section empirically evaluates the proposed \modelname on three real-world datasets. The goal is to answer the four following research questions (RQs). 
\begin{itemize}[leftmargin=*]
    \item \textbf{RQ1:} Is \modelname effective compared with other multi-task learning algorithms? 
    \item \textbf{RQ2:} Does each designed task help in improving the recommendation performance?
    \item \textbf{RQ3:} Does the designed task focusing and gradient magnitude balancing module plays a role in \modelname? 
    \item \textbf{RQ4:} What is the impact of hyper-parameter $\alpha$ and $\tau$ in \modelname?
    \item \textbf{RQ5:} How does the training curve of the newly proposed multi-task training algorithm look like?
\end{itemize}

\subsection{Experimental Setup}
In this section, we illustrate our experiment setting to justify the effectiveness of \modelname, including datasets, baselines, evaluation method, and detailed experiment setting.

\subsubsection{Datasets} 
We compare \modelname against baselines on three real-world datasets with different sizes. The detailed data statistics of each dataset are shown in Table~\ref{table1}. Epinion~\footnote{\url{https://www.cse.msu.edu/~tangjili/datasetcode/truststudy.htm}} collects user's reviews on items with user's social friends on e-commerce website. Based on the available information, the Social Prediction task (predict whether two users are social friends) and the Category Prediction task (predict whether two items belong to the same category) are constructed as auxiliary tasks.
Video Game and Office~\cite{he2016ups} are both from the Amazon platform. For the Video Game dataset, we build a co-view and co-buy prediction task to predict whether two items are co-viewed/co-buyed together frequently. As for the Office dataset, we construct the Alignment/Uniformity task~\cite{wang2022towards} to directly regularize on the embedding table as well as a rating prediction task to predict the user's explicit rating score towards items. Following previous researches~\cite{he2016ups,he2020lightgcn}, a 5-core setting is applied to all datasets, i.e., we remove users/items with less than $5$ interactions to keep the dataset quality.

\begin{table}[t]
  \caption{Statistics of the Datasets.}
  \label{table1}
  \begin{tabular}{l| c c c}
        \toprule
        \textbf{Dataset} & \textbf{Epinion} & \textbf{Video Game} & \textbf{Office} \\
        \hline
        \textbf{\#Users} & 22,167 & 55,223 & 4,905 \\

        \textbf{\#Items} & 296,278 & 17,408 & 2,420 \\

        \textbf{\#Interactions} & 922,267 & 497,577 & 51,441 \\
        \textbf{Density} & 0.014\% & 0.051\% & 0.433\% \\
        \hline
        \textbf{Task 1} & Social & Co-View & Alignment \\
                        & Prediction & Prediction & Uniformity \\
        \hline
        \textbf{Task 2}  & Category & Co-Buy & Rating \\
                        & Prediction & Prediction & Regression \\
        
        \bottomrule
  \end{tabular}
\end{table}

\subsubsection{Baselines}
We compare \modelname with 7 baselines in multi-task learning to test the effectiveness of \modelname.
\begin{itemize}[leftmargin=*]
    \item EW: It is the most basic method by directly adding all losses together to optimize user/item representation.
    \item GradDrop~\cite{chen2020just}: It detects the gradient conflict on each parameter and drops the conflict signal with a scale-measured probability. A smaller scale is easier to be dropped.
    \item PCGrad~\cite{yu2020gradient}: Project Conflicting Gradients (PCGrad) changes one of the optimization directions of conflict gradients to make the gradients conflict-free.
    \item GradVac~\cite{wang2020gradient}: Gradient Vaccine (GradVac) is the updated version of PCGrad to further align the unconflict gradients.
    \item CAGrad~\cite{liu2021conflict}: Conflict-averse gradient descent (CAGrad) minimizes the average loss function by leveraging the worst local improvement of individual tasks to regularize the trajectory.
    \item Aligned MTL~\cite{senushkin2023independent}: It proposes to use a condition number of a linear system of gradients as a stability criterion to guide the multi-task learning optimization.
    \item Nash MTL~\cite{navon2022multi}: It achieves state-of-the-art performance on several MTL benchmarks. Nash MTL suggests conceptualizing the process of combining gradients as akin to a bargaining game, wherein individual tasks engage in negotiation to mutually agree upon a collective direction for parameter updates.
    \item RLW~\cite{lin2021reasonable}: Random Loss Weight (RLW) assigns loss weight randomly in each epoch to explore the different tasks.
\end{itemize}

\begin{table*}[t]
    \caption{Overall Comparison. The best is bolded, and the runner-up is underlined.}
    \label{tab:overall_comp}
    \centering
    \begin{tabular}{l|l|ccccccccccc}
         \toprule
         Dataset & Metric & EW & GradDrop & PCGrad & GradVac & CAGrad & Aligned MTL & Nash MTL & RLW & \modelname & $Imp$\\
         \hline
   
        \multirow{6}{*}{Epinion} 
        & R@5     & 0.0130 & 0.0131 & 0.0142 & 0.0142 & 0.0142 & 0.0123 & \underline{0.0156} & 0.0151 & \textbf{0.0170} & 8.97\% \\
        & R@10    & 0.0219 & 0.0224 & 0.0233 & 0.0233 & 0.0240 & 0.0203 & \underline{0.0251} & 0.0238 & \textbf{0.0266} & 5.97\% \\
        & R@20    & 0.0356 & 0.0355 & 0.0364 & 0.0364 & 0.0379 & 0.0314 & \underline{0.0408} & 0.0381 & \textbf{0.0413} & 1.22\%  \\
        & R@40    & 0.0562 & 0.0570 & 0.0573 & 0.0573 & 0.0596 & 0.0494 & \underline{0.0633} & 0.0598 & \textbf{0.0639} & 0.94\%  \\ \cline{2-12}
        & N@5     & 0.0128 & 0.0130 & 0.0137 & 0.0137 & 0.0139 & 0.0121 & \underline{0.0153} & 0.0146 & \textbf{0.0168} & 9.80\% \\
        & N@10    & 0.0161 & 0.0164 & 0.0170 & 0.0170 & 0.0175 & 0.0150 & \underline{0.0188} & 0.0178 & \textbf{0.0202} & 7.44\% \\
        & N@20    & 0.0205 & 0.0207 & 0.0213 & 0.0213 & 0.0221 & 0.0186 & \underline{0.0239} & 0.0225 & \textbf{0.0249} & 4.18\%  \\
        & N@40    & 0.0263 & 0.0267 & 0.0272 & 0.0272 & 0.0282 & 0.0236 & \underline{0.0303} & 0.0286 & \textbf{0.0312} &  2.97\% \\
        \bottomrule
        \bottomrule

        \multirow{6}{*}{Video Game} 
        & R@5     & 0.0435 & 0.0435 & 0.0408 & 0.0511 & 0.0397 & 0.0321 & 0.0454 & \underline{0.0461} & \textbf{0.0467} & 1.30\% \\
        & R@10    & 0.0716 & 0.0716 & 0.0654 & 0.0657 & 0.0643 & 0.0528 & \underline{0.0732} & 0.0730 & \textbf{0.0750} & 2.45\% \\
        & R@20   & 0.1109 & 0.0974 & 0.1044 & 0.1046 & 0.1006 & 0.0843 & 0.1128 & \underline{0.1129} & \textbf{0.1154} & 2.21\%  \\
        & R@40   & \underline{0.1677} & 0.1480 & 0.1591 & 0.1599 & 0.1533 & 0.1296 & \underline{0.1682} & 0.1668 & \textbf{0.1713} & 1.84\%  \\ \cline{2-12}
        & N@5     & 0.0347 & 0.0347 & 0.0326 & 0.0329 & 0.0313 & 0.0256 & 0.0360& \underline{0.0365} & \textbf{0.0373} & 2.19\% \\
        & N@10    & 0.0451 & 0.0451 & 0.0416 & 0.0419 & 0.0404 & 0.0333 & 0.0463 & \underline{0.0464} & \textbf{0.0478} & 3.01\% \\
        & N@20   & 0.0571 & 0.0497 & 0.0535 & 0.0538 & 0.0515 & 0.0429 & 0.0584 & \underline{0.0585} & \textbf{0.0601} & 2.73\% \\
        & N@40   & 0.0712 & 0.0624 & 0.0670 & 0.0676 & 0.0646 & 0.0541 & 0.0722 & \underline{0.0719} & \textbf{0.0740} &  2.92\% \\
        \bottomrule
        \bottomrule

        \multirow{6}{*}{Office} 
        & R@5     & 0.0243 & 0.0281 & 0.0213 & 0.0213 & 0.0325 & 0.0144 & 0.0224 & \underline{0.0336} & \textbf{0.0347} & 3.27\% \\
        & R@10    & 0.0401 & 0.0458 & 0.0350 & 0.0349 & 0.0513 & 0.0248 & 0.0403 & \underline{0.0553} & \textbf{0.0594} & 7.41\% \\
        & R@20   & 0.0655 & 0.0752 & 0.0576 & 0.0576 & 0.0819 & 0.0417 & 0.0685 & \underline{0.0849} & \textbf{0.0905} &  6.59\% \\
        & R@40   & 0.1074 & 0.1139 & 0.0893 & 0.0891 & 0.1247 & 0.0676 & 0.1100 & \underline{0.1321} & \textbf{0.1474} &  11.58\% \\ \cline{2-12}
        & N@5     & 0.0192 & 0.0229 & 0.0165 & 0.0165 & 0.0260 & 0.0124 & 0.0198 & \underline{0.0259} & \textbf{0.0281} & 8.49\% \\
        & N@10    & 0.0253 & 0.0295 & 0.0217 & 0.0217 & 0.0332 & 0.0163 & 0.0267 & \underline{0.0344} & \textbf{0.0372} & 8.13\% \\
        & N@20   & 0.0334 & 0.0387 & 0.0289 & 0.0289 & 0.0428 & 0.0217 & 0.0362 & \underline{0.0439} & \textbf{0.0473} &  7.74\% \\
        & N@40   & 0.0443 & 0.0489 & 0.0372 & 0.0372 & 0.0539 & 0.0287 & 0.0474 & \underline{0.0562} & \textbf{0.0619} &  10.14\% \\
        \bottomrule 
    \end{tabular}
\end{table*}

\subsubsection{Evaluation Method} 
We randomly split the dataset into a training set (60\%), validation set (20\%), and test set (20\%). Models are trained on the training set, hyper-parameters are tuned based on the performance on the validation set, and the reported results are from the test set.
We test the effectiveness of \modelname on Top-K personalized recommendation task with three widely adopted evaluation metrics: Recall@K (R@K), Hit Ratio@K (HR@K), and Normalized Discounted Cumulative Gain@K (N@K). To comprehensively assess our model, we present evaluation results for Top-20 and Top-40 recommendations in our experiments.
Our evaluation approach calculates average metrics across all users within the test set. These metrics are computed based on the rankings of items that users have not interacted with. To align with established research practices~\cite{he2020lightgcn,wang2022towards}, we adopt the complete full ranking for evaluation. This method entails ranking all items that a user has not yet engaged with, thereby providing a comprehensive evaluation.

\subsubsection{Experiment Settings} 
To make a fair comparison, we test \modelname and all the other baselines under the same setting. User/Item embedding size $d$ is fixed to 32 with Xavier~\cite{Glorot2010Xavier} initialization. To keep simplicity, we directly use the embedding table as the Encoder. It is to be noted that \modelname also supports other encoders. We fix the batch size as 2048 and utilize Adam optimizer~\cite{kingma2014adam} to optimize the parameters. For the remaining hyper-parameters, we used the grid search method to find the optimal settings for \modelname: the learning rate is searched in \{0.1,0.05,0.01,0.005,0.001\}, coefficient of weight decay is tuned in \{$1e^{-2}, 1e^{-4}, 1e^{-6}, 1e^{-8}$\}. We search the task focusing base $\alpha$ from 1.0 to 1.2 with a step size of 0.02, and the temperature $\tau$ in gradient magnitude balancing module in \{$10,1,1e^{-1}, 1e^{-2}, 1e^{-3}, 1e^{-4},1e^{-5},1e^{-6}$\}. We stopped the training early without improvement in successive 10 epochs on the main recommendation task for all the experiments and reported the results on the test set.

\subsection{RQ1: Performance Comparison}
We compare \modelname on 3 real-world datasets with 8 baselines in multi-task learning. The experiment results are shown in Table~\ref{tab:overall_comp}. The best performance is bolded, and the second best is underlined, respectively. We can have the following observations:
\begin{itemize}[leftmargin=*]
    \item \modelname consistently outperforms the second-best method on all three datasets. On the Epinion dataset, \modelname even surpasses the second best with $9.8\%$ improvement in N@5. It justifies the effectiveness of \modelname compared with all previous multi-task training algorithms for RecSys.
    \item The improvement of \modelname for different datasets varies. \modelname achieves remarkable improvement on Epinion and Office datasets while only achieving about $2\%$ improvement on the Video Game dataset. We assume it is because the designed tasks do not provide much extra information. The co-view, co-buy item prediction task can be easily inferred from user-item interactions.
    \item RLW nearly always ranks the second best across all three datasets. RLW randomly assigns task weights in each epoch during training. It is a quite simple method to add more randomness for gradient exploration. It surpasses the sophisticated designed gradient manipulation methods such as PCGrad and GradVac by a large margin. The previous gradient manipulation methods directly consider all the parameter sets without personalized consideration, which hinders their performance.
    \item EW is still a strong baseline for the personalized RecSys task. EW directly adds the losses from different tasks together, which is the simplest multi-task learning method. On the Video Game dataset, it even ranked the second best on R@40 and HR@40. It indicates we do not need a more complicated scheme like previous algorithms. At the same time, \modelname surpasses EW by a large margin, which is also simple with heuristic designs.
\end{itemize}

\begin{figure}[htbp]
      \begin{center}
        \includegraphics[width=0.46\textwidth]{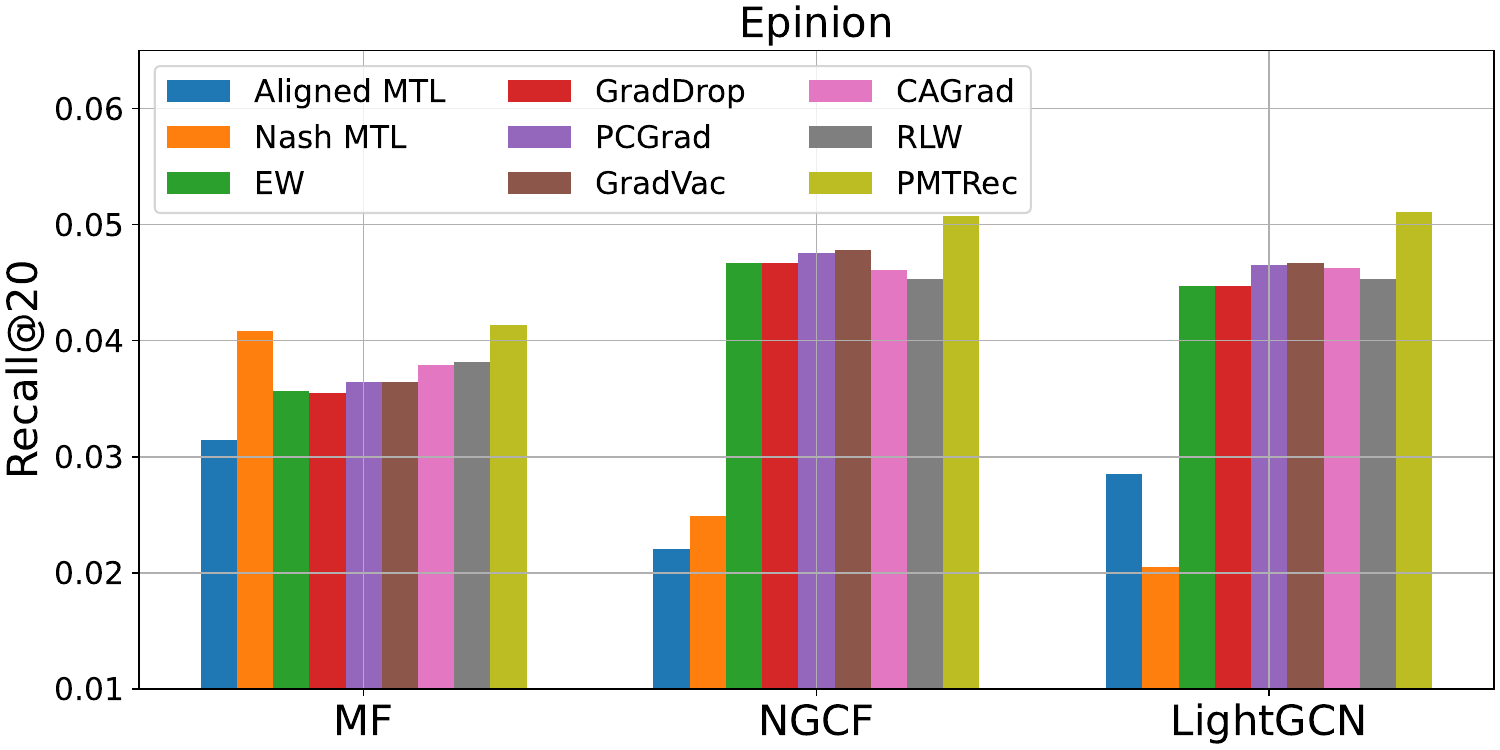}
      \end{center}
        \caption{Experiments on different encoders}
        \label{encoder}
\end{figure}

To show the compatibility of \modelname with the current widely used RecSys, we further conduct experiments on different encoders, including MF~\cite{rendle2012bpr}, NGCF~\cite{he2017neural} and LightGCN~\cite{he2020lightgcn}. Experiment results are shown in Figure~\ref{encoder}. We can observe that \modelname always achieves the best performance regardless of the encoder. It shows that \modelname is compatible with current RecSys algorithms. Besides, we can also observe the performance improvement with a more advanced encoder. It shows the improvement brought by multi-task training is orthogonal to the encoder, and we can integrate it within current RecSys algorithms to achieve better performance.

\begin{table*}[t]
\caption{Task Investigation. Bolded scores indicate the best performance, while underlined scores represent the second-best.}
\label{tab:comparison}
\begin{tabular}{l|l|cccc|cccc|cccc}
\toprule

\multirow{2}{*}{Model} & \multirow{2}{*}{Tasks} & \multicolumn{4}{c}{Epinion} & \multicolumn{4}{c}{Video Game} & \multicolumn{4}{c}{Office} \\

\cmidrule(r){3-6} \cmidrule(r){7-10} \cmidrule(r){11-14}
& & R@5 & R@20 & N@5 & N@20 & R@5 & R@20 & N@5 & N@20 & R@5 & R@20 & N@5 & N@20 \\
\midrule
\multirow{4}{*}{MF} & BPR & 0.0117 & 0.0349 & 0.0118 & 0.0198 & 0.0427 & 0.1085 & 0.0336 & 0.0553 & 0.0281 & 0.0729 & 0.0243 & 0.0402 \\
& \quad +T1 & 0.0158 & 0.0395 & 0.0158 & 0.0238 & \textbf{0.0469} & 0.1151 & 0.0373 & 0.0600 & 0.0372 & 0.0860 & 0.0304 & 0.0470\\
& \quad +T2 & 0.0167 & \textbf{0.0414}& 0.0166 & 0.0249 & 0.0465 & \textbf{0.1161} & 0.0368 & 0.0596 & 0.0232 & 0.0684 & 0.0190 & 0.0345 \\
& \quad +T1,T2 & \textbf{0.0170} & \underline{0.0413} & \textbf{0.0168} & \textbf{0.0250} & \underline{0.0467} & \underline{0.1154} & \textbf{0.0374} & \textbf{0.0601} & \underline{0.0347} & \textbf{0.0905} & \underline{0.0281} & \textbf{0.0473} \\ \bottomrule
\bottomrule

\multirow{4}{*}{NGCF} & BPR & 0.0118 & 0.0294 & 0.0121 & 0.0180 & 0.0433 & 0.1105 & 0.0344 & 0.0565 & 0.0232 & 0.0689 & 0.0197 & 0.0355 \\ 
& \quad +T1 & 0.0181 & 0.0461 & 0.0185 & 0.0278 & 0.0463 & 0.1139 & 0.0369 & 0.0593 & 0.0281 & 0.0812 & 0.0201 & 0.0382 \\
& \quad +T2 & 0.0184 & 0.0481 & 0.0187 & 0.0287 & 0.0450 & 0.1117 & 0.0359 & 0.0579 & 0.0202 & 0.0688 & 0.0145 & 0.0308 \\
& \quad +T1,T2 & \textbf{0.0195} & \textbf{0.0507} & \textbf{0.0199} & \textbf{0.0302} & \textbf{0.0463} & \textbf{0.1138} & \textbf{0.0370} & \textbf{0.0594} & \textbf{0.0354} & \textbf{0.0823} & \textbf{0.0295} & \textbf{0.0454} \\ \bottomrule
\bottomrule

\multirow{4}{*}{LightGCN} & BPR & 0.0122 & 0.0320 & 0.0129 & 0.0195 &  0.0418 & 0.1041 & 0.0331 & 0.0538 & 0.0260 & 0.0723 & 0.0216 & 0.0380 \\
& \quad +T1 & 0.0173 & 0.0459 & 0.0178 & 0.0273 & 0.0442 & 0.1133 & 0.0355 & 0.0585 & 0.0338 & 0.0846 & 0.0258 & 0.0450 \\
& \quad +T2 & \textbf{0.0201} & 0.0499 & 0.0196 & 0.0296 & 0.0445 & 0.1138 & 0.0356 & 0.0585 & 0.0221 & 0.0713 & 0.0165 & 0.0332 \\
& \quad +T1,T2 & \underline{0.0198} & \textbf{0.0511} & \textbf{0.0197} & \textbf{0.0301} & \textbf{0.0448} & \textbf{0.1144} & \textbf{0.0355} & \textbf{0.0586} & \textbf{0.0338} & \textbf{0.0931} & \textbf{0.0280} & \textbf{0.0466} \\

\bottomrule 
\end{tabular} 
\end{table*}

\subsection{RQ2: Task Investigation}

This section aims to investigate whether each designed task helps improve the recommendation performance. Towards this end, we experiment with individual tasks, and Table~\ref{tab:comparison} shows the influence of tasks on recommendation performance across different datasets. To have a thorough investigation on the effectiveness of multi-task learning in RecSys, we conduct these experiments with different encoders, including MF~\cite{rendle2012bpr}, NGCF~\cite{wang2019neural} and LightGCN~\cite{he2020lightgcn}. We can have the following observations in this experiments:
\begin{itemize}[leftmargin=*]
    \item We observe the improvements across all three datasets comparing pure BPR performance and the one combining BPR, T1, and T2. On the Epinion dataset, \modelname improves the BPR performance from $0.0117$ to $0.0170$ on R@5 with two extra auxiliary tasks, a $45.2\%$ improvement. The huge improvement indicates the necessity of combining knowledge from extra tasks to alleviate the data sparsity problem in RecSys. The experiment results also validate \modelname can effectively fuse the knowledge from auxiliary tasks to improve the RecSys performance.
    \item There is a general trend of performance improvements when additional tasks are integrated. For example, the pure RecSys task on the Epinion dataset achieves an N@5 of 0.0118. Adding T1 increases to 0.0158, and adding T2 increases to 0.0166. Jointly combining the two tasks achieves the best performance of 0.0168. 
    \item The impact of individual tasks varies between datasets. On Epinion and Video Game datasets, RecSys performance gradually improves with more tasks. However, the addition of T2 on the Office dataset results in a performance drop on all three backbones, which indicates the auxiliary tasks can also have a negative impact on the RecSys performance, and it should be designed carefully to align it with the main RecSys task.
    \item We can observe the effectiveness of \modelname on all the three backbones. Nearly \modelname always performs best with both T1 and T2 on all three backbones. It shows the \modelname can be easily integrated into current RecSys algorithms.
\end{itemize}


\subsection{RQ3: Ablation Study}

\begin{figure}[tbp]
      \begin{center}
        \includegraphics[width=.23\textwidth]{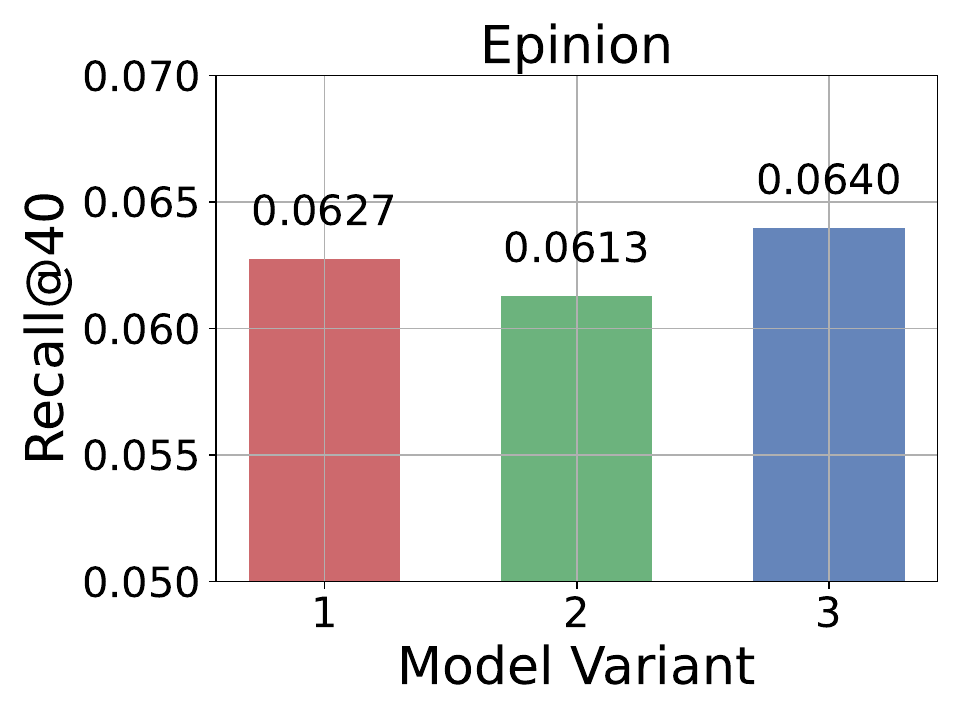}
        \includegraphics[width=.23\textwidth]{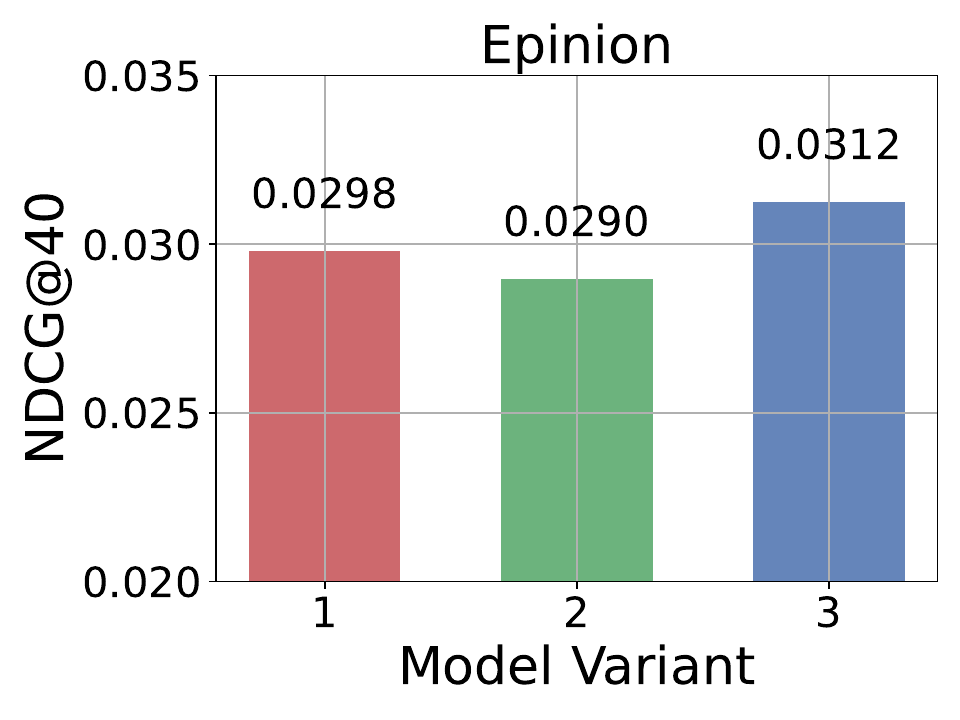}
        \includegraphics[width=.23\textwidth]{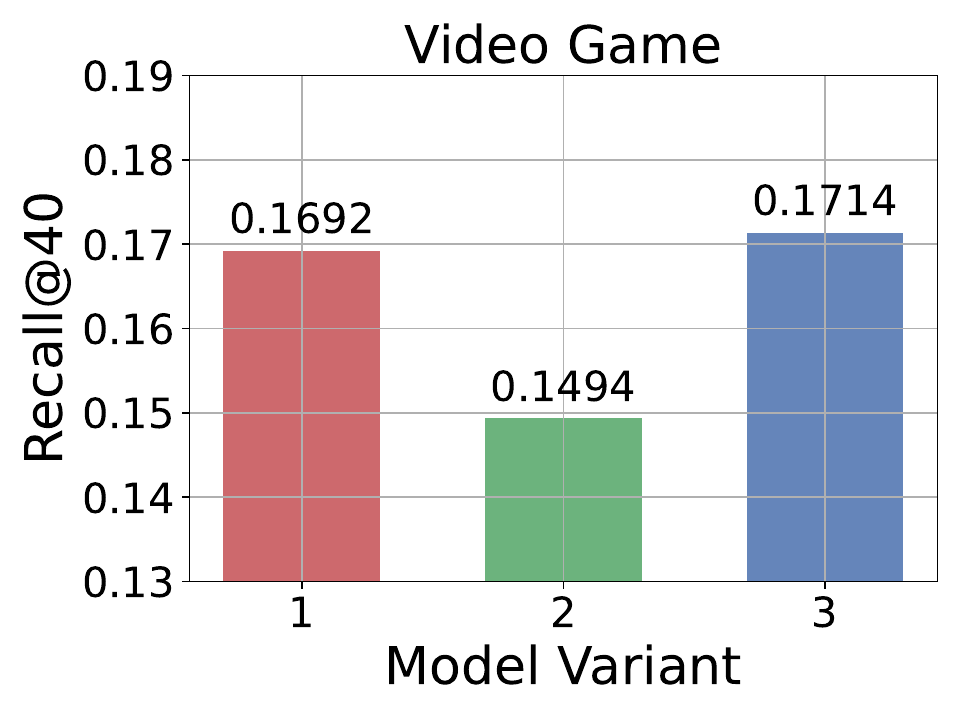}
        \includegraphics[width=.23\textwidth]{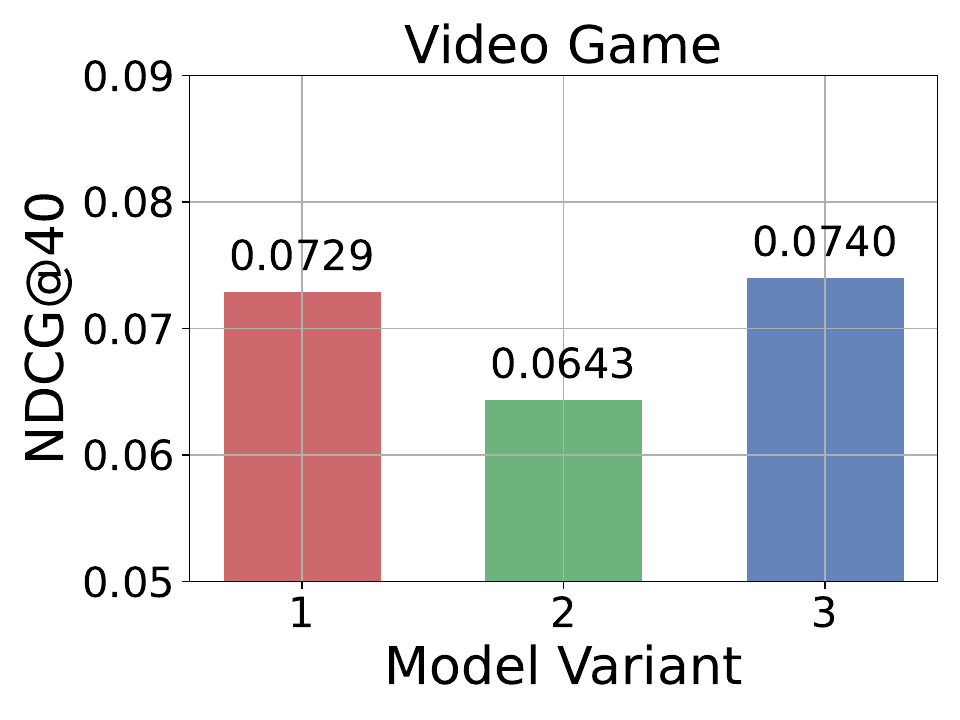}
        \includegraphics[width=.23\textwidth]{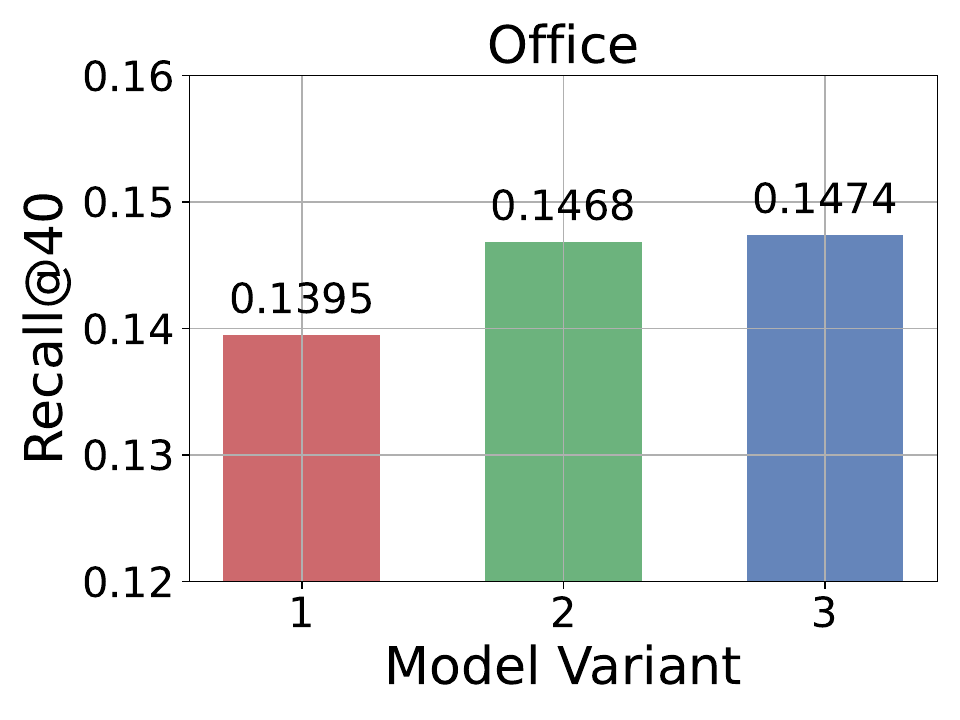}
        \includegraphics[width=.23\textwidth]{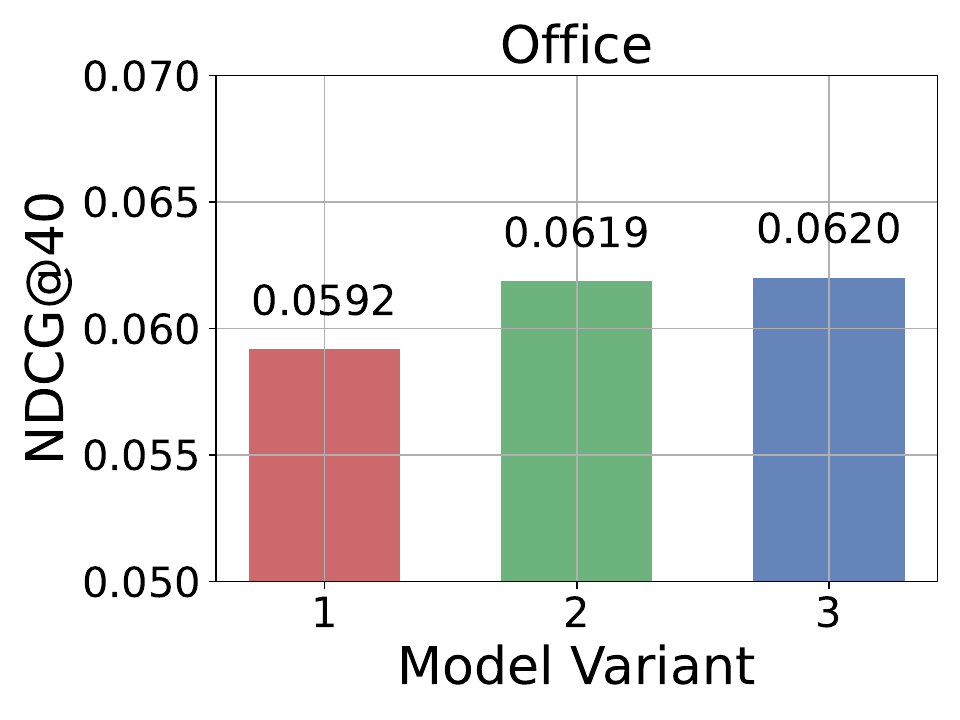}
      \end{center}
        \caption{Ablation study of \modelname. Variant 1 omits the task focusing module, Variant 2 excludes the gradient magnitude balancing module, and Variant 3 represents the full \modelname.
}
        \label{Ablation study}
\end{figure}

To investigate the significance of each module, we conducted an ablation study on the task focusing and gradient magnitude balancing modules, and figure~\ref{Ablation study} displays the corresponding results. From the figure, we can have the following observations:
\begin{itemize}[leftmargin=*]
    \item We observe performance deterioration across all three datasets upon omitting any module. It shows each module contributes to the effectiveness of \modelname.
    \item Excluding the task-focusing module leads to a slight decline in the Epinion and Video Game datasets but substantially decreases the Office dataset. Notably, Recall@40 and NDCG@40 drop by 5.7\% and 4.7\% on Office, respectively. It reveals that gradually focusing the training on the main RecSys task can improve the main RecSys task, which is overlooked by all previous multi-task training algorithms.
    \item The omission of the gradient magnitude balancing module has a marked impact on the Epinion and Video Game datasets. Specifically, the Recall@40 and NDCG@40 metrics are decreased by 4.4\% and 7.6\% on Epinion, and by 14.7\% and 15.1\% on Video Game, highlighting the module's critical role in model performance.
\end{itemize}

\subsection{RQ4: Hyper-Parameter Sensitivity}
\begin{figure}[tbp]
      \begin{center}
        \includegraphics[width=.23\textwidth]{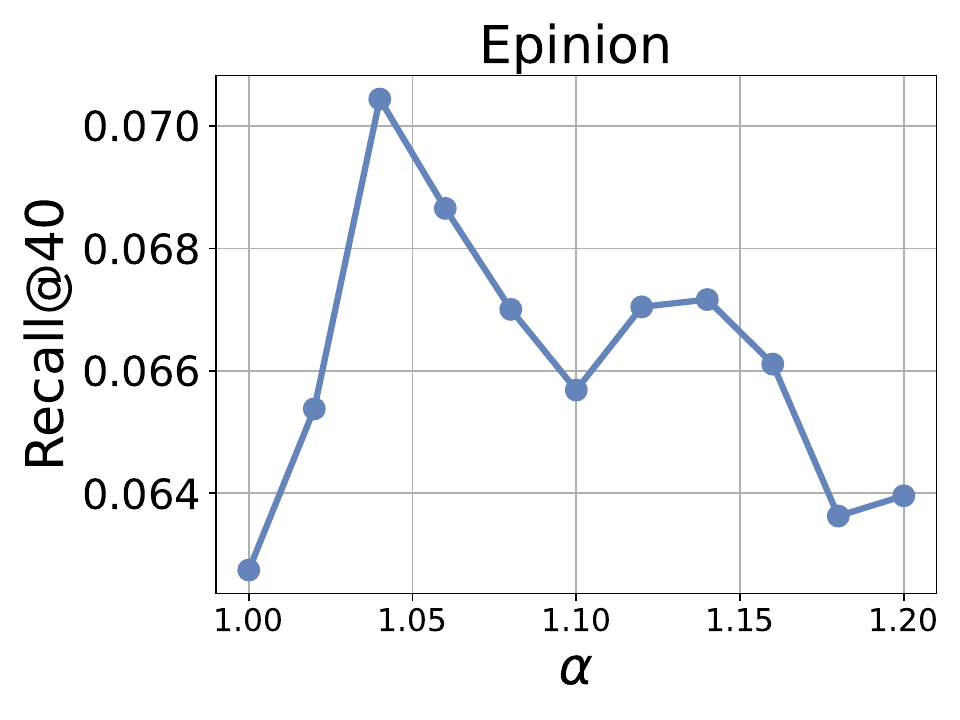}
        \includegraphics[width=.23\textwidth]{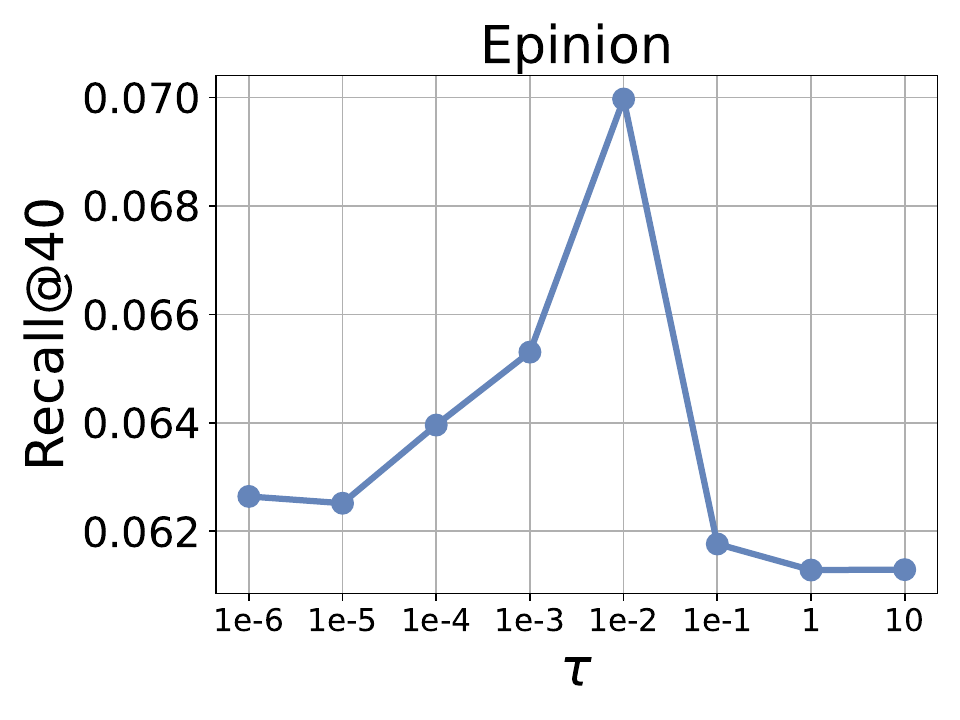}
        \includegraphics[width=.23\textwidth]{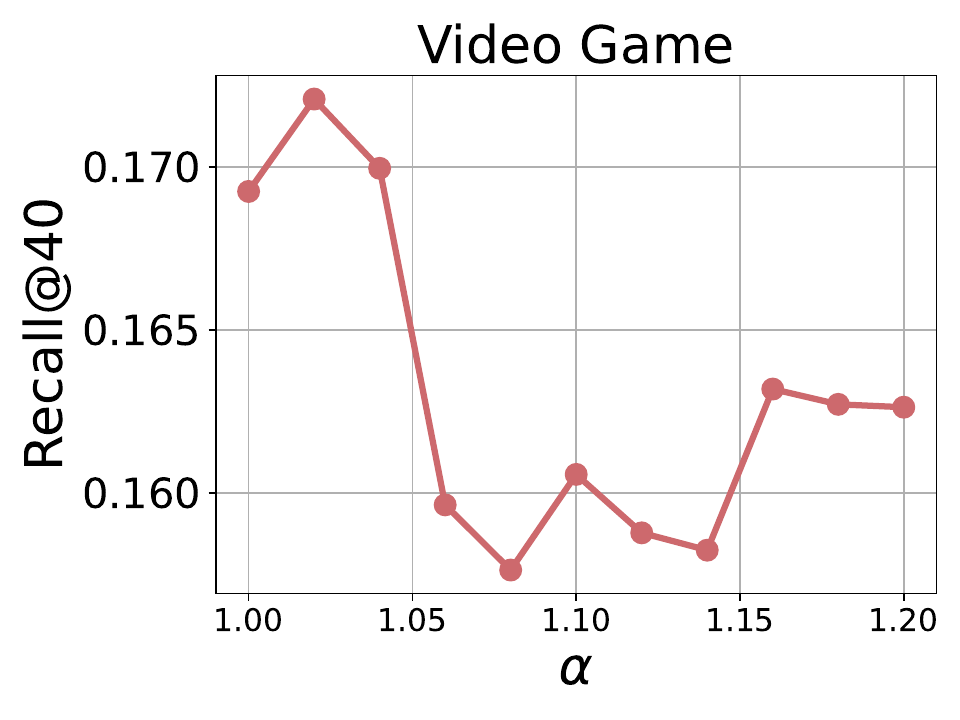}
        \includegraphics[width=.23\textwidth]{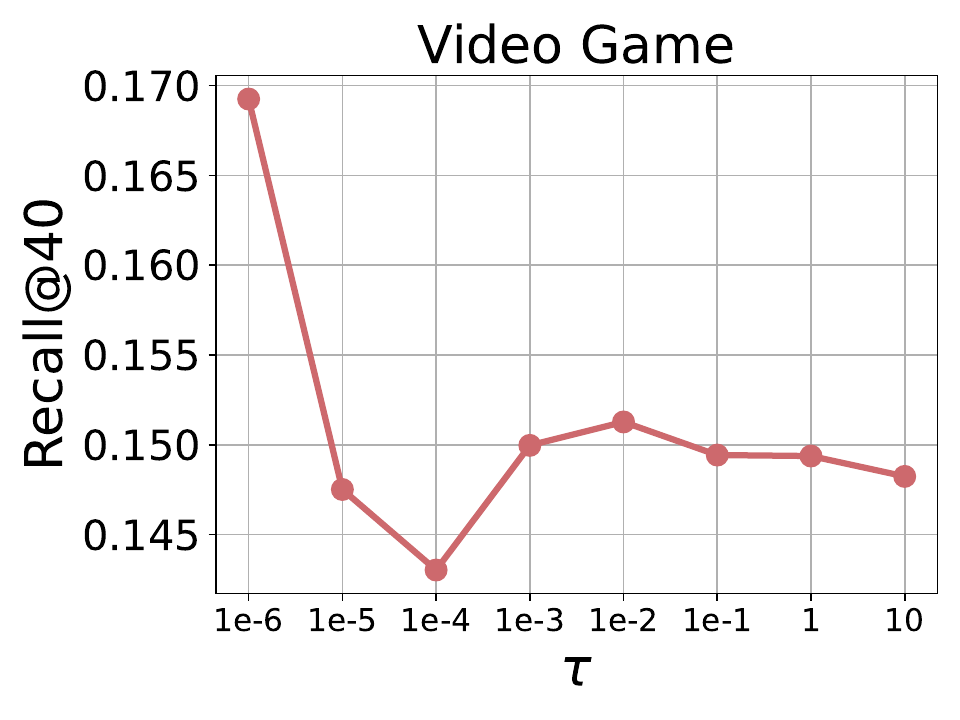}
        \includegraphics[width=.23\textwidth]{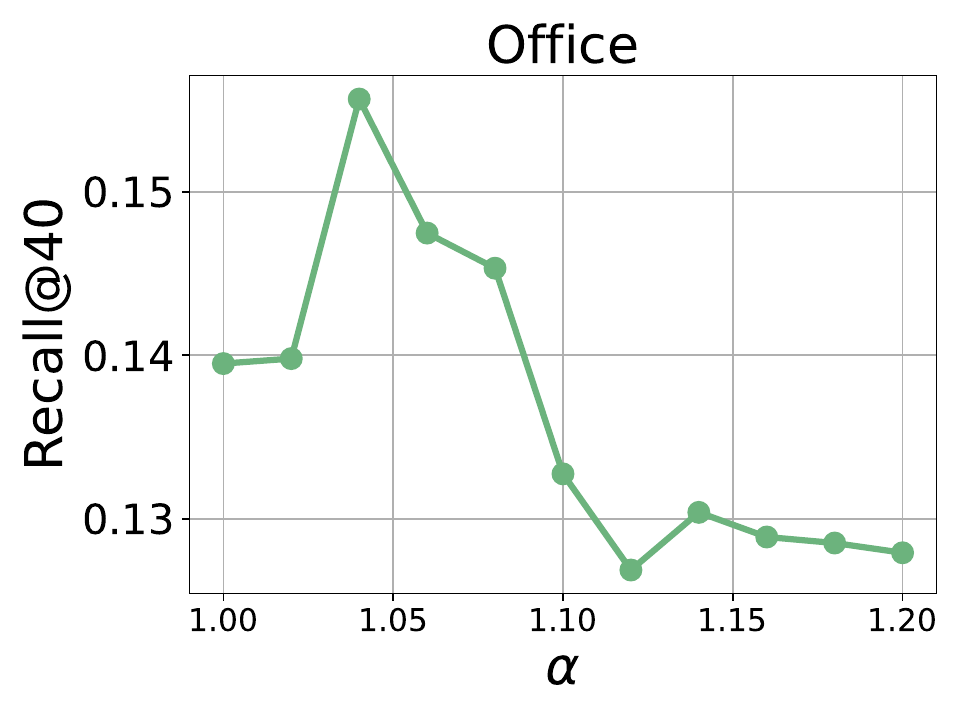}
        \includegraphics[width=.23\textwidth]{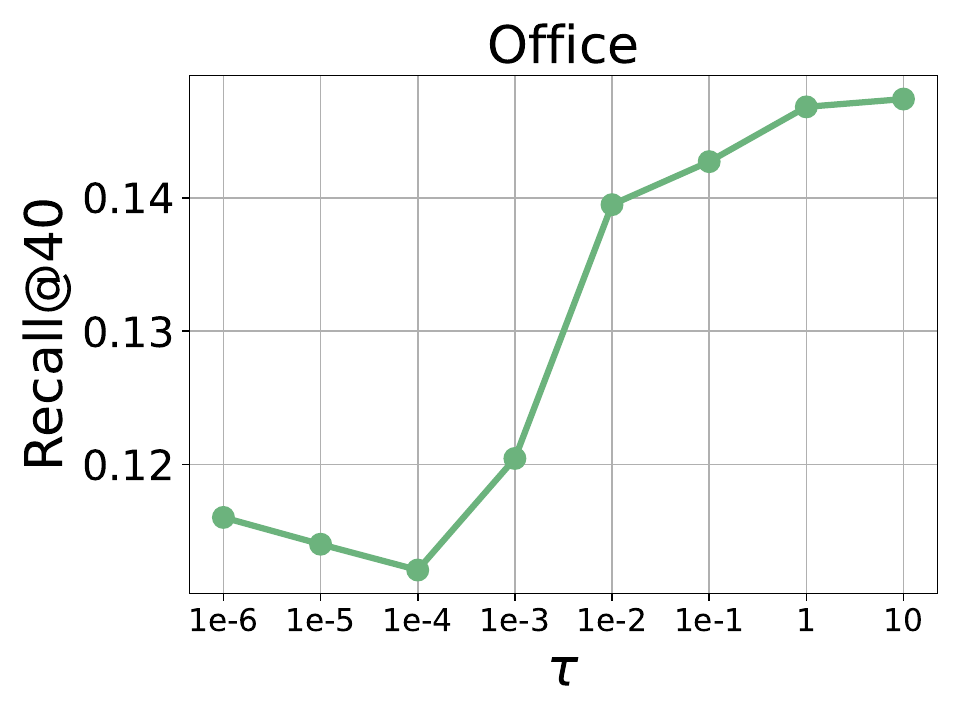}
      \end{center}
        \caption{Sensitivity analysis of \modelname's hyper-parameters: $\alpha$ is the increment base in the task focusing module, while $\tau$ balances the gradient magnitude in the balancing module.}
        \label{Parameter-analysis}
\end{figure}

This section aims to study the Hyper-parameter sensitivity of \modelname. It introduces two hyper-parameters in different modules. $\alpha$ is the multiplying base in the task-focusing module. With a larger $\alpha$, the training of \modelname is faster focused on the main RecSys task. $\tau$ is the temperature in the gradient magnitude balancing module. All task weights will be the same without being impacted by different gradient magnitudes when $\tau \to +\infty$. Conversely, \modelname will only count the gradient with the most prominent norm when $\tau \to -\infty$. We show the experiment results on all three datasets in Figure~\ref{Parameter-analysis}. From the results, we can observe that:
\begin{itemize}[leftmargin=*]
    \item With the increase of $\alpha$, the performance of \modelname firstly increases to the peak point and then quickly decreases. At the initial stage, increasing $\alpha$ will focus the training of \modelname on the RecSys task at the final stage, showing the advantages of task focusing. When $\alpha$ is large, \modelname focuses the training on RecSys quickly without enough exploration of auxiliary tasks. It reveals the importance of learning procedures on auxiliary tasks.
    \item As $\tau$ increases, distinct curve patterns emerge across various datasets. For the Epinion dataset, performance initially reaches a peak before declining. Conversely, optimal performance in the Video Game dataset is achieved with the smallest $\tau$, while the Office dataset attains its highest performance with the largest $\tau$. Auxiliary tasks are constructed in different ways for different datasets. It shows uniqueness of datasets and task construction.
    \item In most cases, \modelname does not perform best when $\alpha=1$ or $\tau=1$. When the hyper-parameter is equal to 1, the corresponding module does not play a role, leading to inferior performance. It re-validates the effectiveness of each designed module in \modelname.
\end{itemize}


\subsection{RQ5: Training Curve Comparison}

\begin{figure}[tbp]
      \begin{center}
        \includegraphics[width=0.44\textwidth]{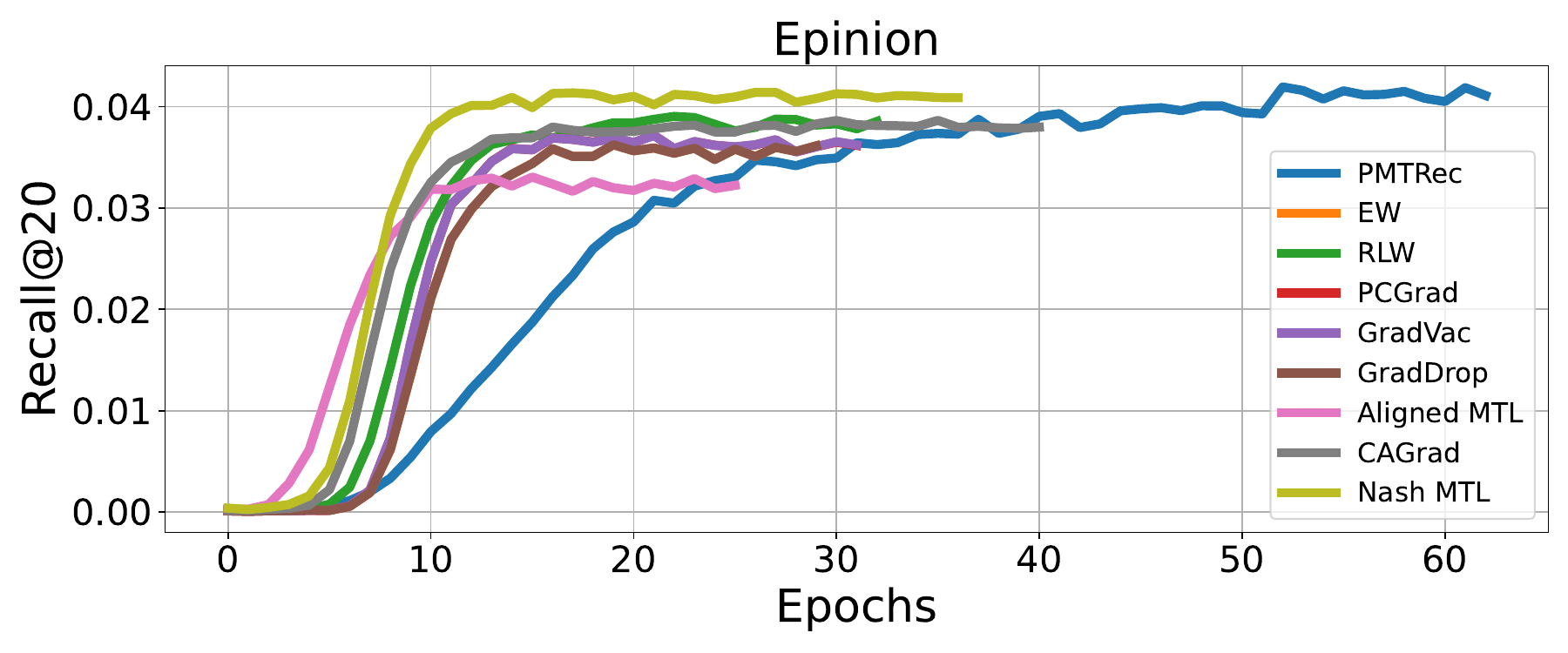}
        \includegraphics[width=0.44\textwidth]{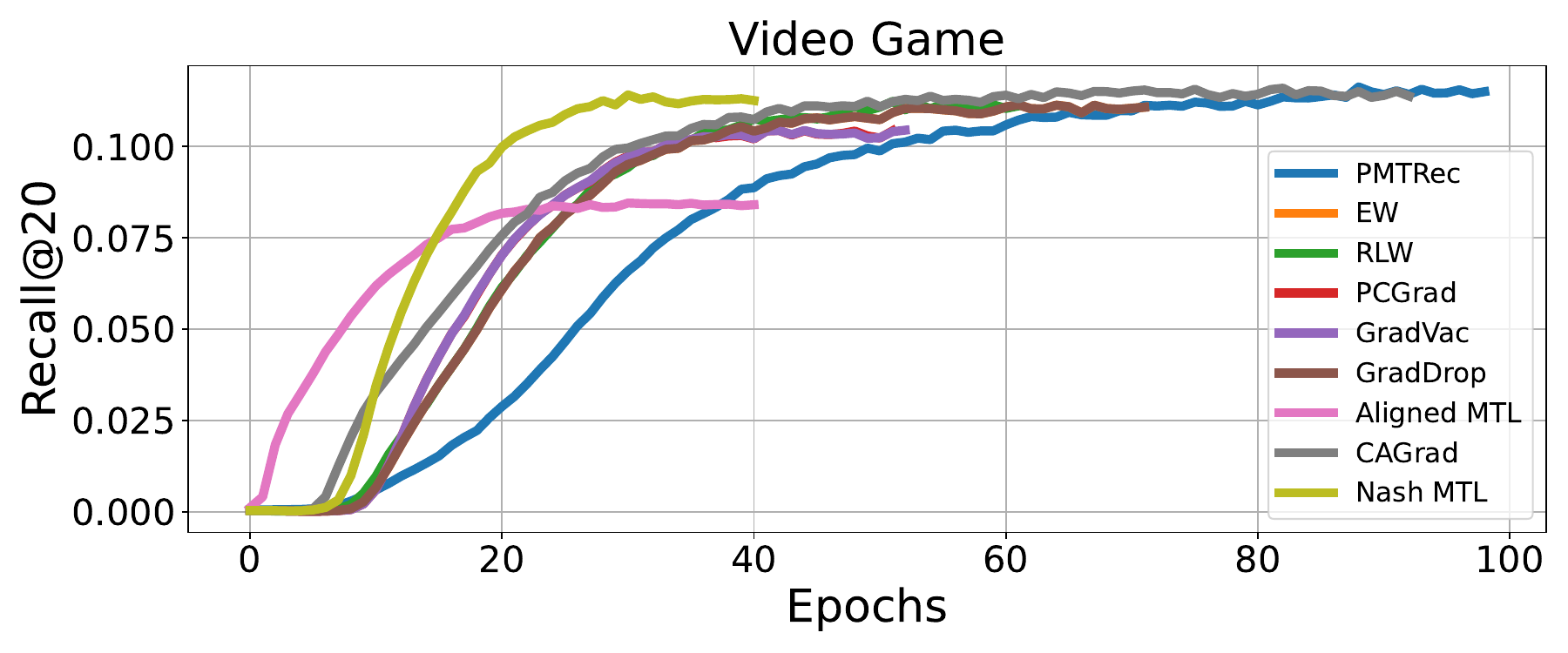}
        \includegraphics[width=0.44\textwidth]{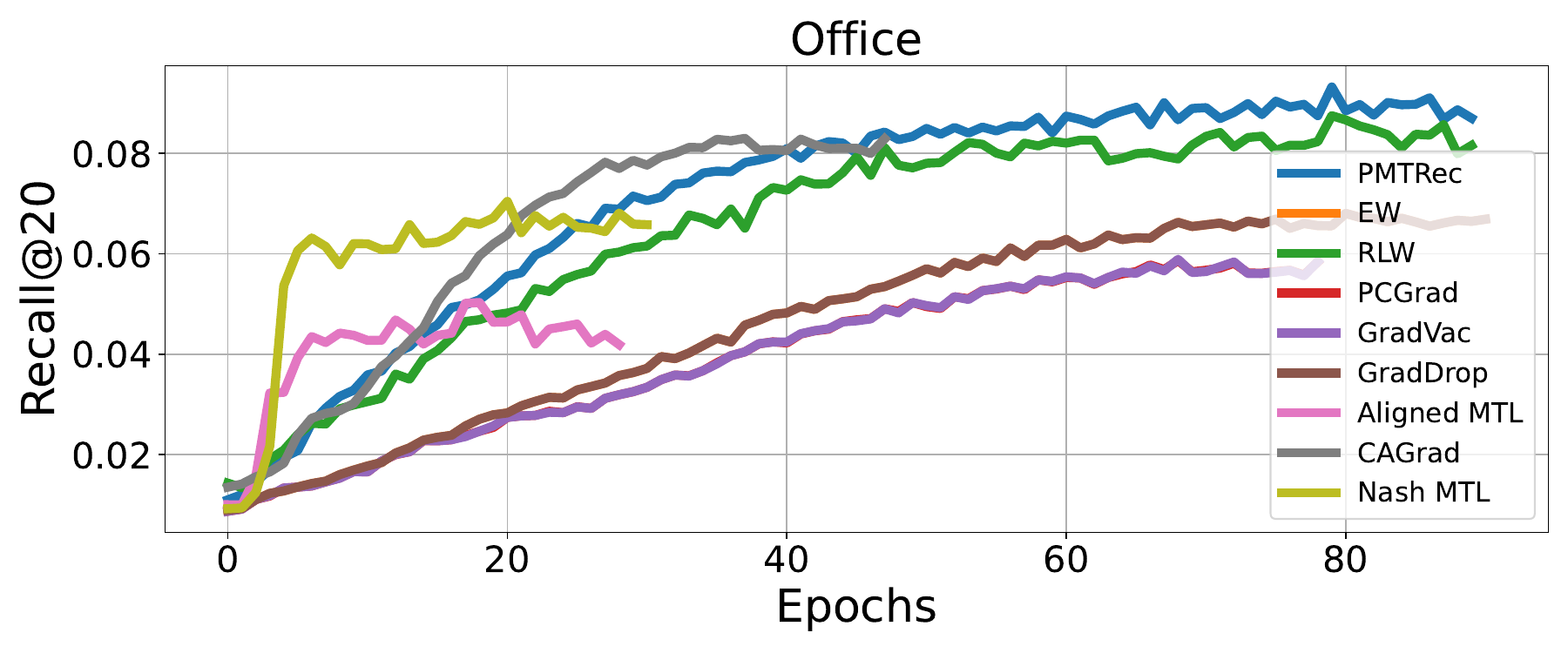}
      \end{center}
        \caption{The training curve on three datasets.}
        \label{training}
\end{figure}

As a new multi-task training algorithm, we further show the training curve of \modelname in Figure~\ref{training}. All the baselines are also included for comparison. We can have the following observations:
\begin{itemize}[leftmargin=*]
    \item On the Epinion dataset, other baselines improve quickly at the first few epochs and reach stable on inferior performance. The improvement of \modelname is more stable and achieves the highest performance at the final stage. It is because, at the initial stage, \modelname tends to explore auxiliary tasks first without fast improvement on the RecSys task. With more epochs, the task-focusing module gradually plays a role in enforcing model train toward the RecSys task. Thus, we observe a gradual improvement of \modelname on the RecSys task.
    \item Similar trend as Epinion dataset is also observed on the Video Game dataset. \modelname explores the auxiliary task first and improves slowly and steadily at the beginning stage. After sufficient exploration, it achieves the best performance at the final stage.
    \item On Office dataset, \modelname also improves stability and achieves the highest performance compared to other multi-task training methods. The training curve on the Office dataset justifies the effectiveness of \modelname.
    \item Another interesting finding is \modelname seems more stable across datasets. For all three datasets, \modelname improves at similar paces. They all improve fastest on 25 epochs and reach the best performance on about 60 epochs. However, other baselines do not seem to have a similar property. All the baselines on the Epinion dataset improve fast at the initial stage, while no similar trend is observed on the Office dataset. We assume personalized updates in \modelname lead to this observation as it captures the personalization character of the RecSys task.
\end{itemize}

\section{Related Work}
This section introduces the most related literature with this paper and the difference between current methods and our newly proposed \modelname. It includes Multi-task Learning and Representation Learning in Recommender Systems.
\subsection{Multi-task Learning}
Multi-Task Learning (MTL)~\cite{caruana1997multitask,li2020personality} is an approach that focuses on collectively training multiple interconnected tasks. The goal is to enhance their generalization capabilities by capitalizing on shared insights across these tasks. There has been a fast development of MTL in recent years. 
Current methods can be classified into two categories: conflict resolution-based methods and randomness-based methods.
Conflict resolution-based methods aim to identify and solve the gradient conflict/imbalance from different tasks.
MGDA-UB~\cite{sener2018multi} utilizes multiple gradient descent algorithms to find a common descending direction among all gradients to avoid the gradient conflict by solving a quadratic programming problem.
GradNorm~\cite{chen2018gradnorm} aims to solve the gradient imbalance problem by constraining the gradient magnitudes to be similar.
PCGrad~\cite{yu2020gradient} modifies the optimization direction of conflicting gradients, redirecting them to achieve gradient conflict resolution and promote coherence among the gradients.
GradVac~\cite{wang2020gradient} represents an enhanced iteration of PCGrad, designed to align non-conflicting gradients in similar directions more effectively.
ATTITTUD~\cite{dery2021auxiliary} and Forkmerge~\cite{jiang2024forkmerge} proposes auxiliary task update to judge the effectiveness of primary task and auxiliary tasks.
CAGrad~\cite{liu2021conflict} employs the concept of worst local improvement from individual tasks, and it minimizes the average loss function, thereby guiding the algorithm's trajectory through regularization.
Aligned MTL~\cite{senushkin2023independent} suggests utilizing a conditioned number of linear gradient systems as a stability criterion, guiding optimizing multi-task learning to avoid conflict.
TAWT~\cite{chen2021weighted} proposes target aware cross task weighted training algorithm.
IMTL-G~\cite{liu2021towards} strives to identify an aggregated gradient characterized by uniform-length projections onto the gradients of each individual task.
RotoGrad~\cite{javaloy2021rotograd} addresses conflict issues by aligning gradient magnitudes and directions concurrently. It achieves this by introducing a trainable rotation matrix to modify the gradient direction for each task, followed by the computation of weights that enforce uniform gradient magnitudes across all tasks.
On the other hand, randomness-based methods aim to add more gradient exploration based on multiple tasks.
UW~\cite{kendall2018multi} employs consistent uncertainties as loss weights for individual tasks, which are adaptively adjusted through backpropagation.
GradDrop~\cite{chen2020just} posits that conflict arises from discrepancies in gradient signs across various tasks. To mitigate this conflict, GradDrop drops gradient values based on magnitude-measured probability.
RLW~\cite{lin2021reasonable} introduces variability by assigning loss weights randomly in each epoch, facilitating the exploration of task optimization from diverse perspectives.
Nash MTL~\cite{navon2022multi} attains top-notch performance across various Multi-Task Learning (MTL) benchmarks. The approach advocates for interpreting the gradient combination process as a form of bargaining game, where individual tasks negotiate to reach a consensus on a shared direction for parameter updates.

All the previous methods do not consider personalized gradient combination, which stands \modelname out from previous methods. It is the first multi-task training algorithm designed for RecSys that supports assigning different task weights for each user/item.

\subsection{Representation Learning for RecSys}
The core of existing recommender systems is learning user/item representations. 
BPR-MF~\cite{rendle2012bpr} is one of the most popular methods to characterize user-item interactions and train user/item representations. 
FM~\cite{rendle2010factorization} further incorporates real-value attributes of users and items to learn representations.
Additionally, if given a user-item rating matrix, NMF~\cite{lee2000algorithms} is widely adopted to learn user/item embeddings. 
As deep learning technology develops, NCF~\cite{he2017neural} is proposed to encode user/item IDs with neural networks as embeddings. And NFM~\cite{he2017neural} enables the feature incorporation in representation learning with neural networks.
Recently, the successes of graph neural network~{GNN}~\cite{kipf2016semi,wu2019simplifying,velivckovic2017graph} prompt the graph-based collaborative filtering, such as NGCF~\cite{wang2019neural} and LightGCN~\cite{he2020lightgcn}. 
Those graph-based collaborative filtering methods learn user/item representation via aggregating neighbor information, which characterizes the high-order signals in the user-item graph.
Learning user/item representation based on graph embedding also opens more potential to harness additional information. 
GraphRec~\cite{fan2019graph} and ConsisRec~\cite{yang2021consisrec} devise additional graph propagation on social graphs, which enables the joint learning of social and recommendation information.
Besides social information between users, those multiple relations between items~\cite{xu2020knowledge,xu2020product,fan2023zero}, such as co-view and co-buy, also benefit the representation learning for the recommendation.
The success of contrastive learning in representation learning~\cite{wang2020understanding} motivates the novel DirectAU~\cite{wang2022towards} loss for the recommendation.
DirectAU optimizes the user/item representation by minimizing uniformity and alignment losses, eliminating the inefficient negative sampling process in learning representation.
GraphAU~\cite{yang2023graph} further considers the graph-related high-order alignment signal during optimizing user/item representation.
3MN~\cite{zhang20233mn} builds different meta networks for multi-scenario task learning.
With the emerging ability of large language models~(LLM), most recent researches also investigate improving user/item representation with pre-trained large language models for the RecSys task.
P5~\cite{P5} and OpenP5~\cite{xu2024openp5} train different tasks within large language models by next-token-prediction task.
LlmRec~\cite{wei2024llmrec} augments user-item bipartite graph with LLMs to learn a more informative user/item representation.
RLMRec~\cite{ren2023representation} proposes contrastive alignment and generative alignment to align user/item embedding to the large language model encoded space.
CARec~\cite{wangcollaborative} directly aligns the user/item's ID embedding with semantic embedding in the iterative alignment phase.
Previous research shows that a more discriminative, comprehensive, and informative user/item representation is the foundation for effective RecSys.

This paper proposes to directly learn user/item representation under a multi-task learning paradigm and proposes the first personalized multi-task training method \modelname. 
It can encode information from multiple tasks to obtain a more comprehensive and informative user/item representation.

\section{Conclusion}

This paper introduces \modelname, a novel multi-task training algorithm specifically designed for recommender systems. This algorithm distinctively supports personalized gradient combinations for individual user/item. It intricately navigates the backpropagation process, manipulating gradients directly at their level. Within \modelname, two principal modules are constructed: a task-focusing and a gradient magnitude balancing module. The former progressively concentrates training efforts on the primary recommender system (RecSys) task, whereas the latter equilibrates gradients of varying magnitudes. Empirical evaluations conducted on real-world datasets substantiate \modelname's efficacy. As the first multi-task training algorithm designed explicitly for the recommender system, we also open-sourced \modelname for further research.

\bibliographystyle{ACM-Reference-Format}
\bibliography{sample-base}

\appendix

\end{document}